\documentclass[journal=nanoletters,manuscript=letter]{achemso}
\usepackage[latin9,utf8]{inputenc}
\usepackage{float}
\usepackage{textcomp}
\usepackage{amsmath,amssymb,amscd,hyperref}
\usepackage{graphicx}
\makeatletter
\usepackage{url}
\usepackage{grffile}
\usepackage{bbold}
\usepackage{comment}
\usepackage{dirtytalk}
\usepackage{chngcntr}
\usepackage{tabularx}
\usepackage{multirow}
\usepackage{soul}
\usepackage{cleveref}
\usepackage{braket}
\DeclareUnicodeCharacter{2212}{-}
\DeclareUnicodeCharacter{0301}{\hspace{-1ex}\'{ }}

\usepackage{dcolumn}
\usepackage{color}
\DeclareGraphicsExtensions{.png .jpg .pdf}
\usepackage{hyperref}
\hypersetup{
     colorlinks = true,
     linkcolor = blue,
     anchorcolor = blue,
     citecolor = blue,
     filecolor = blue,
     urlcolor = blue
     }
\usepackage{braket}
\usepackage{physics}
\usepackage{array}
\usepackage{booktabs}
\usepackage{soul}

\makeatother

\author{Ze-Hua Tao}
\affiliation{Department of Physics and NANOlight Center of Excellence, University of Antwerp, Groenenborgerlaan 171, 2020 Antwerp, Belgium}

\author{Icaro R. Lavor}
\email{icaro@fisica.ufc.br}
\affiliation{Instituto Federal de Educação, Ciência e Tecnologia do Rio Grande do Norte, Mossoró, Rio Grande do Norte, Brazil}
\alsoaffiliation{Department of Physics and NANOlight Center of Excellence, University of Antwerp, Groenenborgerlaan 171, 2020 Antwerp, Belgium}
\alsoaffiliation{Departamento de F\'{i}sica, Universidade Federal do Ceará, Caixa Postal
6030, Campus do Pici, 60455-900 Fortaleza, Ceará, Brazil}

\author{Hai-Ming Dong}
\email{hmdong@cumt.edu.cn}
\affiliation{School of Materials and Physics, China University of Mining and Technology, Xuzhou 221116, P. R. China}

\author{Andrey Chaves}
\affiliation{Departamento de F\'{i}sica, Universidade Federal do Ceará, Caixa Postal
6030, Campus do Pici, 60455-900 Fortaleza, Ceará, Brazil}
\alsoaffiliation{Department of Physics and NANOlight Center of Excellence, University of Antwerp, Groenenborgerlaan 171, 2020 Antwerp, Belgium}

\author{David Neilson}
\affiliation{Department of Physics and NANOlight Center of Excellence, University of Antwerp, Groenenborgerlaan 171, 2020 Antwerp, Belgium}

\author{Milorad V. Milo\v{s}evi\'{c}}
\email{milorad.milosevic@uantwerpen.be}
\affiliation{Department of Physics and NANOlight Center of Excellence, University of Antwerp, Groenenborgerlaan 171, 2020 Antwerp, Belgium}

\title{Chiral propagation of plasmon polaritons due to competing anisotropies in a twisted photonic heterostructure}

\date{\today}
\begin{document}

\date{\today}
\begin{abstract}
We demonstrate chiral propagation of plasmon polaritons and show it is more efficient and easier to control than the recently observed chiral shear phonon polaritons. We consider plasmon polaritons created in an anisotropic two-dimensional (2D) material, twisted with respect to an anisotropic substrate, to best exploit the competition between anisotropic electron-electron interactions and the anisotropic electronic structure of the host material. Gate voltage and twist angle are then used for precise control of the chiral plasmon polaritons, overcoming the existing restrictions with chiral phonon polaritons. These findings open up feasible opportunities for efficient and tunable plasmon-based nanophotonics and compact high-performance on-chip optical devices.

\begin{figure}[t]
\centering{}\includegraphics[width=0.6\columnwidth]{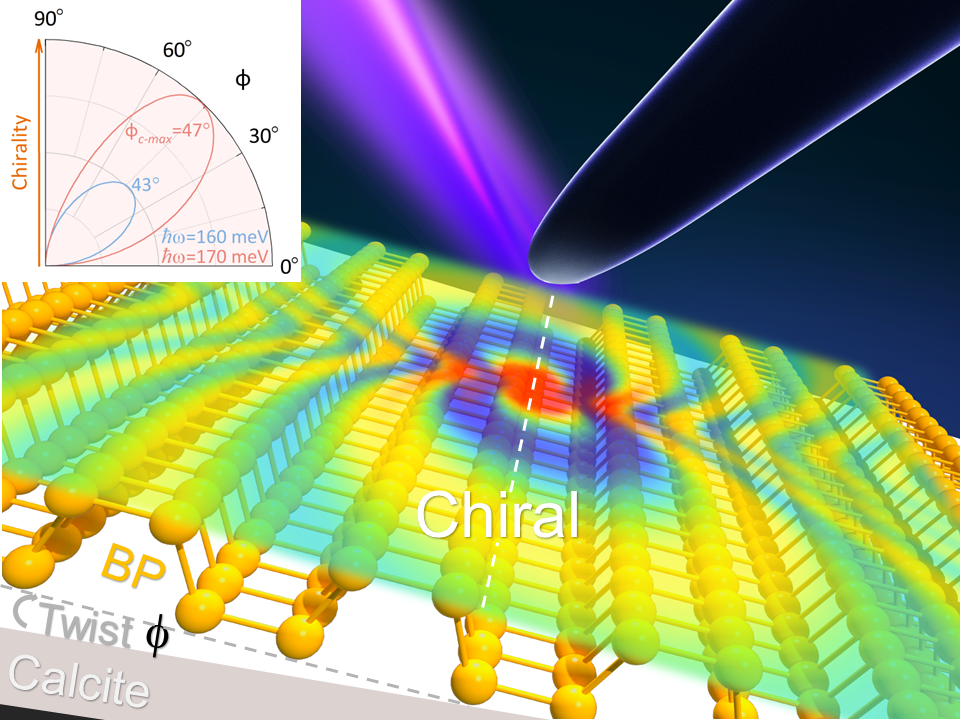}
\end{figure}

\end{abstract}
\maketitle

The precise manipulation of photons at the nanoscale, crucial for the development of nanophotonic devices, has been at the forefront of scientific research in the recent decades~\cite{hu2023nanophotonic,oh2021nanophotonic,chaudhary2019nanophotonic}. 
In this field, polaritons, which are quasiparticles resulting from interactions between photons and material excitations, provide a promising carrier platform to achieve the nanoscale control of light~\cite{wu2022manipulating}. These hybrid excitations hold importance for many applications, such as nanophotonics~\cite{zhang2021interface}, on-chip optics~\cite{liu2016onchip}, and optical integrated devices~\cite{schwarz2014monolithically}. However, directional in-plane polariton propagation remains a challenge due to symmetry considerations, since pronouncedly directional polariton waves require strong anisotropic materials or metasurfaces for their existence~\cite{fu_manipulating_2024}. 

Conversely, anisotropic optical materials, such as black phosphorus (BP)~\cite{low2014plasmons}, capable of supporting plasmons, collective oscillations of the two-dimensional electron liquid~\cite{pogna2024bp}, are promising building blocks for photonic and optoelectronic devices due to their low structural symmetry and significant in-plane optical anisotropy. Furthermore, reducing the crystal symmetry enhances the directional properties of optical phonons and the consequent anisotropic response, facilitating the anisotropic phonon polaritonic phenomena of particular recent interest~\cite{galiffi2023review_aniso_PhPs}.

In this context, a novel polariton with highly directional and asymmetric chiral propagation, stemming from the shear phonon polaritons in low-symmetry natural monoclinic crystals \cite{passler2022shear,hu2023shear,matson2023shear,alvarez2024unidirectional}, has been recently experimentally achieved, opening new possibilities to engineer and control polaritons in van der Waals heterostructures (vdWhs). However, due to the challenging complexity of controlling lattice vibrations~\cite{fu_manipulating_2024}, most phonon polaritons cannot be readily manipulated by optoelectronic methods, which presents significant roadblocks in the fabrication of photonic devices based on these quasiparticles.

\begin{figure}[H]
\centering{}\includegraphics[width=0.6\linewidth]{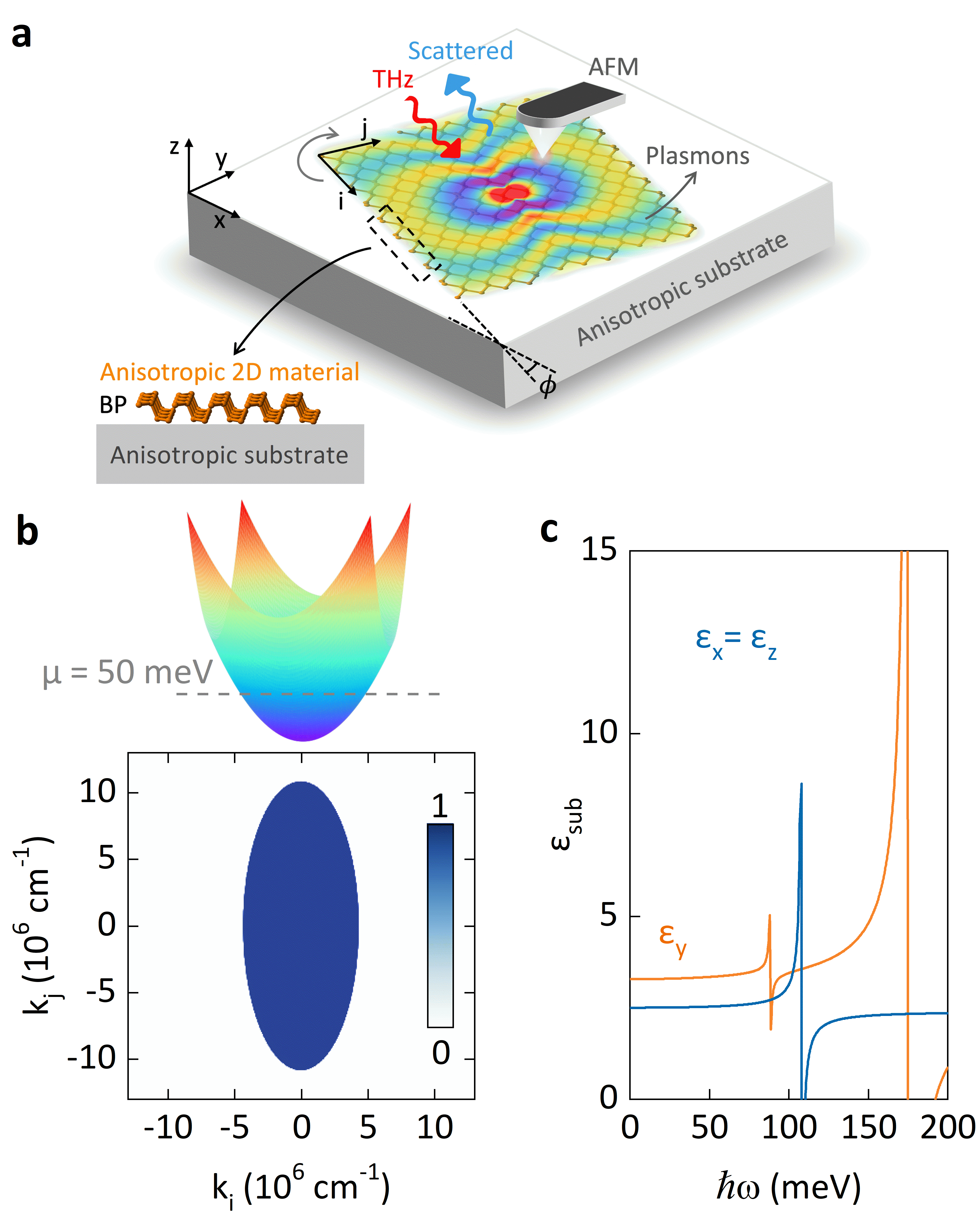}
\caption{\textbf{The designed twisted photonic heterostructure and the origin of the two competing anisotropies.} (a) Schematic illustration of a scatter-type scanning near-field optical microscopy setup, containing an anisotropic 2D material, such as black phosphorus, twisted with angle $\phi$ with respect to an anisotropic substrate, such as calcite. Here $\{i,j\}$ and $\{x,y,z\}$ mark spatial Cartesian coordinates for the 2D material and the 3D substrate, respectively. (b) The anisotropic energy band of a low-energy electron in a black phosphorus monolayer~\cite{qiao_2014,ghosh2017anisotropic} and the corresponding anisotropic Fermi-Dirac distribution $f[E(\boldsymbol{k})]$ (bottom panel) at $T = 0$ K, for chemical potential $\mu$ = 50 meV. (c) The elements of the dielectric tensor $\tilde{\varepsilon}$ for a
calcite substrate as a function of the excitation energy $\hbar\omega$.}
\label{Fig: setup}
\end{figure}

Instead, we reveal in this letter a new type of tunable plasmon polaritons in vdWhs with simultaneously asymmetric and chiral propagation, fostered by competition between two co-existing anisotropies: (i) one arising from the electron-electron interactions between the material that supports plasmon polaritons and the anisotropic substrate; and (ii) the other one due to the anisotropic electronic structure of the plasmonic material itself, such as BP~\cite{low2014graphene}, for example. We realize the latter conditions in a plasmonic system composed of an anisotropic 2D layer (AN-2DL) on top of an anisotropic substrate, where a twist angle between them enables tunability, as illustrated in Figure~\ref{Fig: setup}(a). Experimentally, a terahertz (THz) laser can be used to excite chiral plasmons in an AN-2DL, represented by the background color in Figure~\ref{Fig: setup}(a), which in turn interact with the surrounding environment media~\cite{lundeberg2017tuning}. 
Although we present a generic theoretical concept, as a realistic example from an experimental perspective we have considered BP~\cite{low2014plasmons} as the AN-2DL (exploiting its anisotropic electronic structure as shown in Figure~\ref{Fig: setup}(b)) and calcite, as the anisotropic substrate~\cite{hossain2009electronic} 
(providing the excitation energy-dependent anisotropic dielectric tensor as shown in Figure~\ref{Fig: setup}(c)), unless stated otherwise. The energy dispersion of electrons in this system is obtained from an effective mass approximation for the 2D anisotropic material, while the random phase approximation (RPA) is employed to calculate the chiral plasmonic dispersion in the long-wavelength limit. With this, we first discuss the effect of the competition between the anisotropies present in the plasmonic material and the substrate. Then, we examine the effect of the twist angle $\phi$ on the plasmon propagation for a fixed doping ($\mu$) and excitation energy ($\hbar\omega$), as well as for a fixed twist angle and the varied doping and excitation energy, all for the system sketched in Figure~\ref{Fig: setup}(a). It is important to clearly distinct the role of the twist in this paper from the earlier concept of an electronic band structure modified by twist in bilayer moir\'e superlattices~\cite{huang2022chiralplasmons}. In our investigation, the twist affects plasmon polaritons via the interplay and competition between the anisotropic electron-electron interactions in the twisted heterostructure and the anisotropic electronic structure of the plasmonic 2D material. Thereby optimized and tailored chiral plasmons, excited by scattering-type scanning near-field optical microscopy (s-SNOM) in such vdWhs, can be probed in future experimental works and may yield important advances in further understanding and possible applications of this novel kind of controlled plasmonic propagation.

Within the random phase approximation (RPA), plasmons, i.e. collective modes of the free electron liquid, are described by a dielectric function given by~\cite{giuliani2008quantum}:
\begin{equation}
{\boldsymbol \epsilon}({\boldsymbol q},\omega)=1-\boldsymbol{V}(q,\theta,\phi)\Pi({\boldsymbol q},\omega),
\label{Eq: RPA}
\end{equation}
where $\boldsymbol{V}(q,\theta,\phi)$ is the 2D Fourier transform of the Coulomb interaction between the electrons and $\Pi({\boldsymbol q},\omega)$ is the 2D polarizability. By solving the electrostatics problem, as discussed in the \textit{Supplementary Materials} (SM), one obtains the anisotropic Fourier transform of the electron-electron Coulomb interaction, introduced by the anisotropic substrate located in the region $z \textless 0$ (see Figure~\ref{Fig: setup}(a)).

In order to obtain a general 2D Fourier transform of the Coulomb interaction that encodes all the aspects of the system, we introduced a direction-dependent wave vector $\boldsymbol{q} = (q,\theta)$, where $\theta$ is the angle between the plasmon wave vector $q$ and the $i$-coordinate (see SM), and a direction-dependent effective dielectric function $\varepsilon_{\rm eff}(\theta,\phi)$. This leads to 
\begin{equation}
    \boldsymbol{V}(q,\theta,\phi) =\frac{4\pi e^2}{\varepsilon_{\rm eff}(\theta,\phi)q},
    \label{eq: e-e interaction}
\end{equation}
where the effective dielectric function is given by 
\begin{equation}
\varepsilon_{\rm eff}(\theta,\phi)=[\varepsilon_{x}\varepsilon_{z}{\rm cos}^2(\theta-\phi)+\varepsilon_{y}\varepsilon_{z}{\rm sin}^2(\theta-\phi)]^{1/2}+\varepsilon_0.   
\label{eq:eff epsilon}
\end{equation}
The contribution of the substrate anisotropy is then introduced via the dielectric tensor, defined as:
\begin{equation}
\tilde{\varepsilon}=\left[\begin{array}{ccc}
\varepsilon_{x} & 0 & 0 \\
0 & \varepsilon_{y} & 0 \\
0 & 0 & \varepsilon_{z}
\end{array}\right].
\label{Eq: dielectric tenso}
\end{equation}
In Eq.~(\ref{eq:eff epsilon}), $\varepsilon_0 = 1$ is the dielectric constant of the vacuum in the $z \textgreater 0$ region. As expected, if the substrate is replaced with a material that has an isotropic dielectric constant $\varepsilon$, Eq.~(\ref{eq: e-e interaction}) leads to $\varepsilon_{\rm eff} \rightarrow \varepsilon+\varepsilon_0$. Consequently, the well-known 2D Coulomb interaction $V(q) = 2\pi e^2 / \varepsilon_{\rm env} q$ is obtained in this limit~\cite{low2014plasmons,Dong_2020}, where $\varepsilon_{\rm env}=(\varepsilon+\varepsilon_0)\big/2$ denotes the dielectric constant of the environment. Also, considering calcite as an anisotropic substrate, since the components of the dielectric tensor of this material depend on the excitation energy as shown in Figure~\ref{Fig: setup}(c), we take the corresponding value of $\varepsilon_{i=x,y,z}$ for each plasmon energy considered in our calculations. 

The electron polarizability function is the usual RPA bubble defined as
\begin{equation}
\Pi(\boldsymbol{q},\omega)=g_s\sum_{\boldsymbol{k}} \frac{f[E(\boldsymbol{k}+ \boldsymbol{q})]- f[E(\boldsymbol{k})]}{E(\boldsymbol{k}+\boldsymbol{q})-E(\boldsymbol{k})+\hbar(\omega +i\gamma)},
\label{Eq: polarizability}
\end{equation}
where $g_s=2$ is the spin degeneracy, $f[E(\boldsymbol{k})]=1/[\exp(E(\boldsymbol{k})/(k_BT))+1]$ is the Fermi-Dirac distribution at temperature $T$, $k_B$ is the Boltzmann constant, and $\gamma$ is a parameter that introduces broadening to the excitation resonances in the system.

The anisotropic energy dispersion of the electrons in the AN-2DL, which plays an important role in Eq.~(\ref{Eq: polarizability}) and in the Fermi-Dirac distribution, is obtained from the effective mass approximation for the conduction band of a 2D anisotropic material, defined as~\cite{Rodin2015Collective}
\begin{equation}
    E({\boldsymbol{k}})=\frac{\hbar^2{k_i}^2}{2m_i}+\frac{\hbar^2{k_j}^2}{2m_j}-\mu.
    \label{Eq: energy}
\end{equation}
In Eq.~(\ref{Eq: energy}), ${\boldsymbol{k}} = (k_i, k_j)$ is the wave-vector of the carriers, $\mu$ is the chemical potential, and $m_{i(j)}$ refers to the anisotropic effective mass in the $i$($j$)-direction in the band structure. These effective masses can be obtained e.g. from density functional theory (DFT) calculations \cite{qiao_2014}. Figure ~\ref{Fig: setup}(b) illustrates the energy dispersion obtained from Eq.~(\ref{Eq: energy}) with $m_i = 0.2$ and $m_j = 1.1$, along with its corresponding Fermi-Dirac function in reciprocal space, showing an anisotropic characteristic, as expected.

Finally, the excitation spectrum of plasmons is determined by the vanishing real part of the dynamical dielectric function, defined by Eq.~(\ref{Eq: RPA}), that is, Re[$\boldsymbol{\epsilon}(\boldsymbol{q},\omega)=0$]. In the low-temperature and long-wavelength limit, this yields~\cite{Rodin2015Collective,pyatkovskiy2016dynamical,silva2017anisotropic},
\begin{equation}
\hbar\omega=\sqrt{\frac{2g_se^2\mu}{\varepsilon_{\rm eff}(\theta,\phi)}\mathit{\Lambda}(\theta) q},
\label{Eq: dispersion_aniso}
\end{equation}
with 
\begin{equation}
\mathit{\Lambda}(\theta)=(m_j/m_i)^{1/2}{\rm cos}^2(\theta)+(m_i/m_j)^{1/2}{\rm sin}^2(\theta).
\label{Eq: aniso_structure}
\end{equation}
Thus, the plasmon dispersion obtained from this procedure is controlled via two anisotropic mechanisms: (i) the orientation factor $\mathit{\Lambda}(\theta)$, given by Eq.~(\ref{Eq: aniso_structure}), which captures the 2D anisotropic electronic structure, and (ii) the direction-dependent effective dielectric function $\varepsilon_{\rm eff}(\theta,\phi)$, given by Eq.~(\ref{eq:eff epsilon}), which introduces the anisotropic electronic Coulomb interactions.

\begin{figure}[t]
\centering{}\includegraphics[width=0.5\columnwidth]{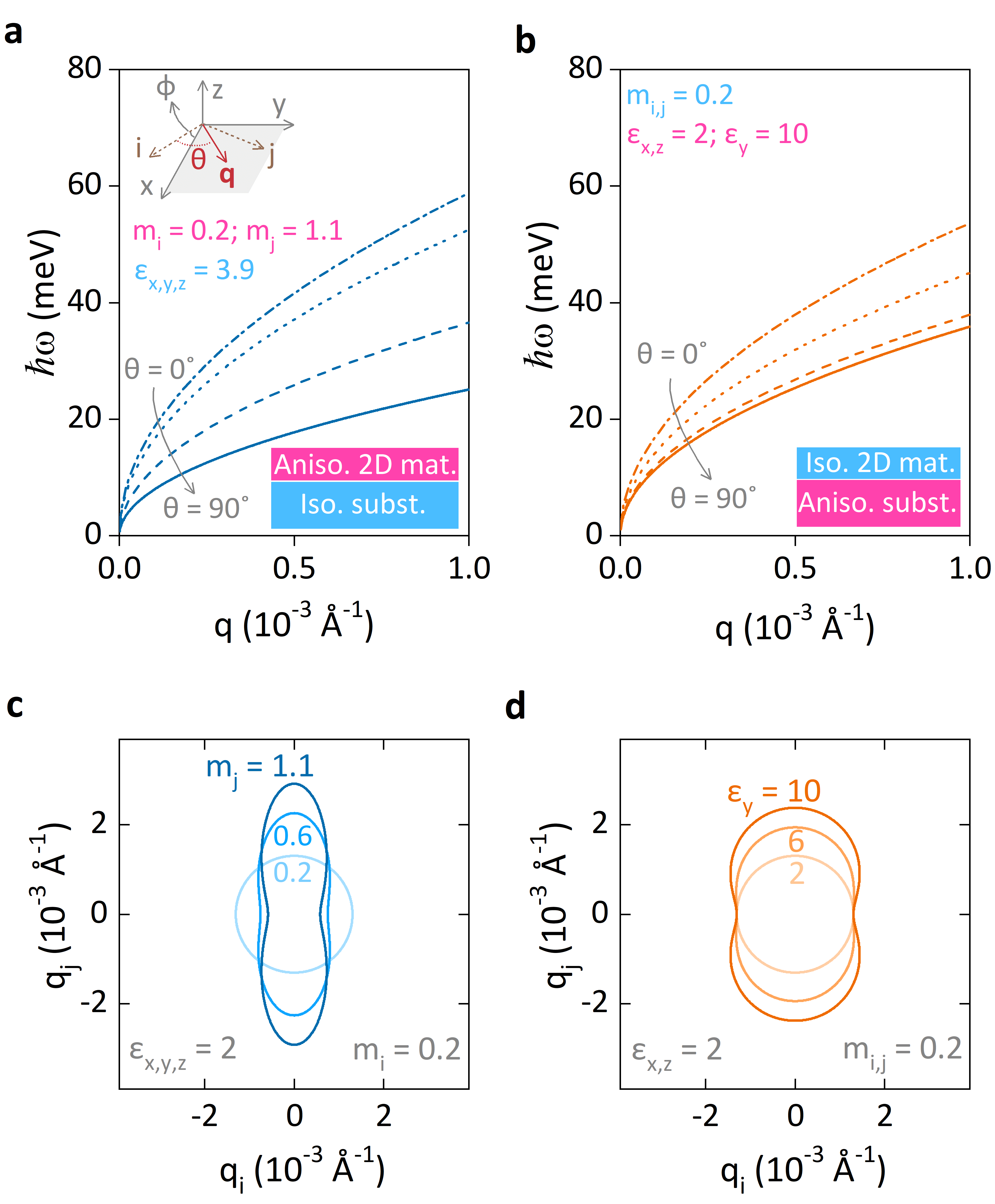}
\caption{\textbf{Dispersions and isofrequency contours of plasmon polaritons, probing the anisotropy of either the 2D material or the substrate.} (a) Plasmon dispersions in directions $\theta = 0$, $30^\circ$, $60^\circ$, $90^\circ$ for an anisotropic BP layer on an isotropic SiO$_2$ substrate. 
(b) Plasmon dispersions in directions $\theta = 0$, $30^\circ$, $60^\circ$, $90^\circ$ for an isotropic 2D layer on an anisotropic substrate. 
(c) Isofrequency contours at a fixed excitation energy $\hbar\omega = 50$ meV for an
anisotropic 2D layer with $m_i = 0.2$, $m_j = 0.2, 0.6, 1.1$, on an isotropic substrate with $\varepsilon = 2$. 
(d) Isofrequency contours for an isotropic 2D layer with $m = 0.2$ on an anisotropic substrate with $\varepsilon_x = \varepsilon_z = 2$, $\varepsilon_y = 2, 6, 10$. In all calculations, the chemical potential was kept fixed at $\mu = 100$ meV, and $\{i,j\}$ and $\{x,y\}$ coordinates coincided ($\phi=0^\circ$ in the inset of panel (a)).}
\label{Fig: dispersion}
\end{figure}

To understand the plasmonic response as impacted by the two anisotropies of the system, one arising from the Coulomb interaction and the other from the polarizability, described by Eqs.~(\ref{eq: e-e interaction}) and (\ref{Eq: polarizability}), respectively, we first separately considered the contributions of the anisotropy of either the 2D material or the substrate, while keeping the other constituent of the heterostructure isotropic. Nevertheless, in all cases we consider plasmon propagation under different angles $\theta$ that define the direction of the wavevector $q$, see inset in Figure~\ref{Fig: dispersion}(a).

The plasmon dispersion of an AN-2DL, with $m_{i(j)} = 0.2$ ($1.1$), on an isotropic substrate with $\varepsilon_{x,y,z} = 3.9$ in the dielectric tensor of Eq.~(\ref{Eq: dielectric tenso}), is shown in Figure~\ref{Fig: dispersion}(a). As one can see, along the $i$-axis, for a fixed plasmon energy, the plasmon wavelength ($\lambda = 2 \pi/q$) is larger when the propagation is along the direction of the smaller mass $m_i$, since the screening is suppressed along this direction when compared to $j$-direction. In other words, the propagation along the $i$-axis leads to a higher plasmon energy. It is important to mention that the parametric values considered here correspond to a realistic situation of a BP monolayer placed on top of the SiO$_2$ substrate.
 
On the other hand, Figure~\ref{Fig: dispersion}(b) presents the opposite situation: an isotropic 2D material is placed on top of an anisotropic substrate. Similarly to the previous case, the plasmon excitation energy is higher when the plasmon propagation direction is along the $x$-axis, since the dielectric constant is smaller in this direction, making Coulomb screening less effective.

Figures \ref{Fig: dispersion}(c) and (d) show the isofrequency contours for a fixed plasmon excitation energy of $\hbar \omega = 50$~meV ($\approx 12$~THz), with contributions of the two anisotropies shown separately (while maintaining $\phi=0^\circ$, i.e. $\{i,j\}$ and $\{x,y\}$ coordinates in the inset of Figure \ref{Fig: dispersion}(a) coincide). Figure~\ref{Fig: dispersion}(c) presents the results for an AN-2DL on top of an isotropic substrate ($\varepsilon_{x,y,z} = 2$), with a fixed mass of $m_i = 0.2$ along the $i$-axis and different values for $m_j$ ranging from 0.2 to 1.1. On the other hand, Figure~\ref{Fig: dispersion}(d) shows the results for an isotropic 2D material ($m_{i,j} = 0.2$) on top of an anisotropic substrate, with $\varepsilon_{x,z} = 2$ and $\varepsilon_y$ ranging from 2 to 10. One observes that both anisotropies, separately, lead to anisotropic isofrequency contours, but with different natures of origin. The former, i.e., the anisotropic electron mass (electronic structure), stretches the isofrequency contours in the direction of increasing effective mass and compresses them in the orthogonal direction, as shown in Figure~\ref{Fig: dispersion}(c), affecting both the $x$ and $y$ directions. This occurs because the electron polarizability incorporates contributions from all the free electrons in the entire space, creating a link between the two directional effects. Conversely, in Figure~\ref{Fig: dispersion}(d), while the dielectric anisotropy of the substrate stretches the contour in the direction of increasing dielectric constant, there is no compression in the orthogonal direction. In this case, the dielectric anisotropy, defined by Eq.~(\ref{eq:eff epsilon}), separates into a contribution only in the $x$-direction for $\theta=0^\circ$ or only in the $y$-direction for $\theta = 90^\circ$. Thus, modifying $\varepsilon_y$ while keeping $\varepsilon_x$ fixed, has no effect on ${\boldsymbol \epsilon}({\boldsymbol q},\omega)$ in the $x$-direction, leaving the screened interaction in the $x$-direction unchanged. Hence a plasmon propagating in the $x$-direction is unaffected.

\begin{figure}[t]
\centering{}\includegraphics[width=0.6\columnwidth]{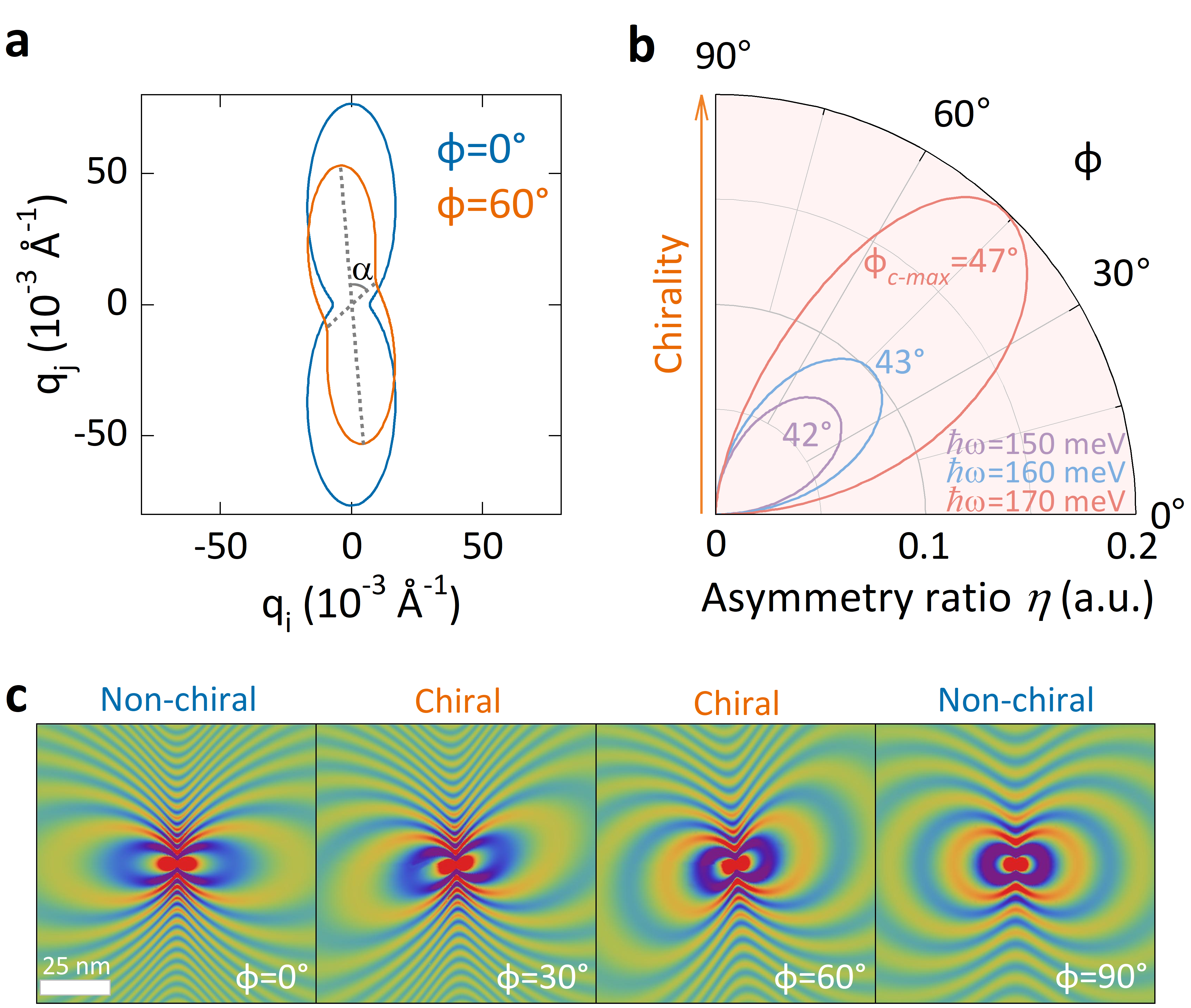}
\caption{\textbf{Anisotropic versus chiral propagation of plasmon-polaritons in a black phosphorous monolayer twisted relative to its calcite substrate.} (a) Isofrequency contours of plasmon polaritons at twist angles $\phi=0^\circ$ and 60$^\circ$ for excitation energy $\hbar\omega = 170$ meV and the chemical potential $\mu = 100$ meV. The dielectric constants of the calcite substrate in this case are $\varepsilon_x = \varepsilon_z = 2.3$ and $\varepsilon_y = 13.2$. The anisotropic electron mass of BP was taken as $m_i = 0.2$, $m_j = 1.1$. (b) The asymmetry ratio $\eta$ for different twist angles $\phi$, for $\hbar\omega = 150$ meV ($\varepsilon_x=\varepsilon_z = 2.2$, $\varepsilon_y=4.8$, $\phi_{c-max} = 42 ^\circ$), $\hbar\omega = 160$ meV ($\varepsilon_x=\varepsilon_z = 2.3$, $\varepsilon_y=6.2$, $\phi_{c-max} = 43 ^\circ$) and 170 meV ($\phi_{c-max} = 47 ^\circ$), all for $\mu = 100$ meV. $\phi_{c-max}$ is the twist angle yielding maximal chirality. (c) Real-space profiles of the s-SNOM generated plasmon polariton for twist angles $\phi=0^\circ$, $30^\circ$, $60^\circ$ and $90^\circ$, for $\hbar\omega = 170$ meV and $\mu = 100$ meV.}
\label{Fig: twist_isofreq}
\end{figure}

Let us now demonstrate that one can indeed generate chirality in plasmonic excitations by exploiting the interplay between the anisotropy in the Coulomb interaction (through $\varepsilon_{\rm eff}(\theta,\phi)$) 
and the anisotropy in the electronic structure (through $m_i\neq m_j$). Chirality generally indicates breaking of specific symmetries~\cite{zhu2018chiralphonons,huang2022chiralplasmons}, which in our system is achieved by the latter interplay of two anisotropies, discussed in more detail in the SM. Figure S1 schematically illustrates how, for twist angles $\phi\neq0$, the mirror symmetry in the 2D layer will be lost when the underlying substrate is anisotropic. The effect will initially become stronger with increasing $\phi$, but weakens as the direction $i$ approaches the direction of the $y$-axis. Figure \ref{Fig: twist_isofreq}(a) shows the isofrequency contours of plasmon polaritons for BP on top of calcite as the excitation-dependent anisotropic substrate~\cite{ma2021ghost} (see SM and Figure~\ref{Fig: setup}(c)), for chemical potential $\mu=100$ meV and excitation energy $\hbar\omega=170$ meV. Without a twist ($\phi=0^\circ$, blue curve in Figure~\ref{Fig: twist_isofreq}(a)), the anisotropy directions for the BP layer and the substrate are aligned, with the enhancement of the plasmon frequency along the same direction, resulting in anisotropic, yet non-chiral, plasmons. In contrast, in presence of a twist, e.g. for $\phi = 60^\circ$, there are \textit{chiral plasmon polaritons}, as seen from their chiral isofrequency contour (orange curve in Figure~\ref{Fig: twist_isofreq}(a)). To quantify the induced chirality, we consider the skew angle $\alpha$ of the isofrequency curve, between the lines connecting the farthest ($q_{\text{max}}$) points and nearest ($q_{\text{min}}$) points to the origin of the isofrequency curve, as shown in Figure~\ref{Fig: twist_isofreq}(a). However, if $q_{\text{max}}$ and $q_{\text{min}}$ are close, even a large skew angle $\alpha$ may not be easily observable in experiment. Therefore, we also account for the anisotropy ratio $\beta$ of the curve, defined as $\beta = (q_{\text{max}} - q_{\text{min}})/q_{\text{max}}$. Finally, we define the asymmetry ratio $\eta=\beta\cdot(\abs{\alpha-90^\circ}/90^\circ)$ as a measure of observable chirality.

As can be seen in Figure~\ref{Fig: twist_isofreq}(a), for $\phi=0^\circ$ the skew angle $\alpha=90^\circ$, and, consequently, asymmetry ratio $\eta$ equals zero. On the other hand, $\alpha \neq 90^\circ$ for $\phi\neq 0^\circ$ or $90^\circ$, which results in the nonzero asymmetry ratio, i.e. chirality is induced. Figure S2 of the SM shows the complete evolution of the chiral plasmon polariton in BP on top of calcite, with increasing the twist angle between the two anisotropies in the system. We observe that the chirality increases with the twist angle until it reaches its maximum at the twist angle dubbed $\phi_{c-max}$, and then decreases beyond this twist angle before completely vanishing at $\phi=90^\circ$. Namely, at 90$^\circ$, the two anisotropic directions of the 2D layer and the substrate are orthogonal, and their combination cannot break the mirror symmetry.

Figure \ref{Fig: twist_isofreq}(b) shows the asymmetry ratio for three values of the excitation energy, $\hbar \omega = 150$, $160$ and $170$ meV, for twist angles $\phi$ ranging from $0^\circ$ to $90^\circ$. The excitation energy $\hbar \omega$ is related to the optical response of the 2D anisotropic layer and it also affects the frequency-dependent dielectric tensor of the anisotropic substrate (cf. Figure~\ref{Fig: setup}(c)). One sees that the twist angle of maximal chirality $\phi_{c-max}$ depends on the plasmon excitation energy $\hbar \omega$ and is highly sensitive to the substrate anisotropy.
The results of Figure \ref{Fig: twist_isofreq}(b) effectively show that maintaining $\varepsilon_x$ constant and increasing $\varepsilon_y$ increases the asymmetry ratio and $\phi_{c-max}$. 
Thus, increasing the substrate anisotropy leads to more pronounced mirror symmetry breaking. These findings indicate that the chiral plasmon polariton can be made experimentally accessible by an appropriate choice of the constituent anisotropic materials and the twist angle close to the $\phi_{c-max}$, according to the prescription shown in Figure~\ref{Fig: twist_isofreq}(b) based on a realistic BP-calcite heterostructure. In Figure S3 of the SM, we provide further theoretical insights into the evolution of chirality when the anisotropies of the 2D material and the substrate are treated as free parameters for a given plasmon excitation energy.

For further facilitated relation to experiment, we show in Figure \ref{Fig: twist_isofreq}(c) the calculated real-space profiles of the asymmetrically propagating plasmon polaritons, as can be experimentally excited with scattering-type scanning near-field optical microscopy (s-SNOM)~\cite{woessner2015ssnom,ma2018hPhPs,ma2021ghost}. The chiral plasmon polariton propagates outward as a radial wave~\cite{woessner2015ssnom} that can be described by a straightforward spherical-wave model $A\simeq{\rm cos}(\boldsymbol{q}\cdot\boldsymbol{r})/\it{r}$ of the geometrical decay of the electric field of the polaritons~\cite{tao2021tailoring,woessner2015ssnom}. These real-space profiles demonstrate once again the clearly asymmetric chiral propagation of the plasmon polaritons in our twisted heterostructure.

\begin{figure}[t]
\centering{}\includegraphics[width=0.6\columnwidth]{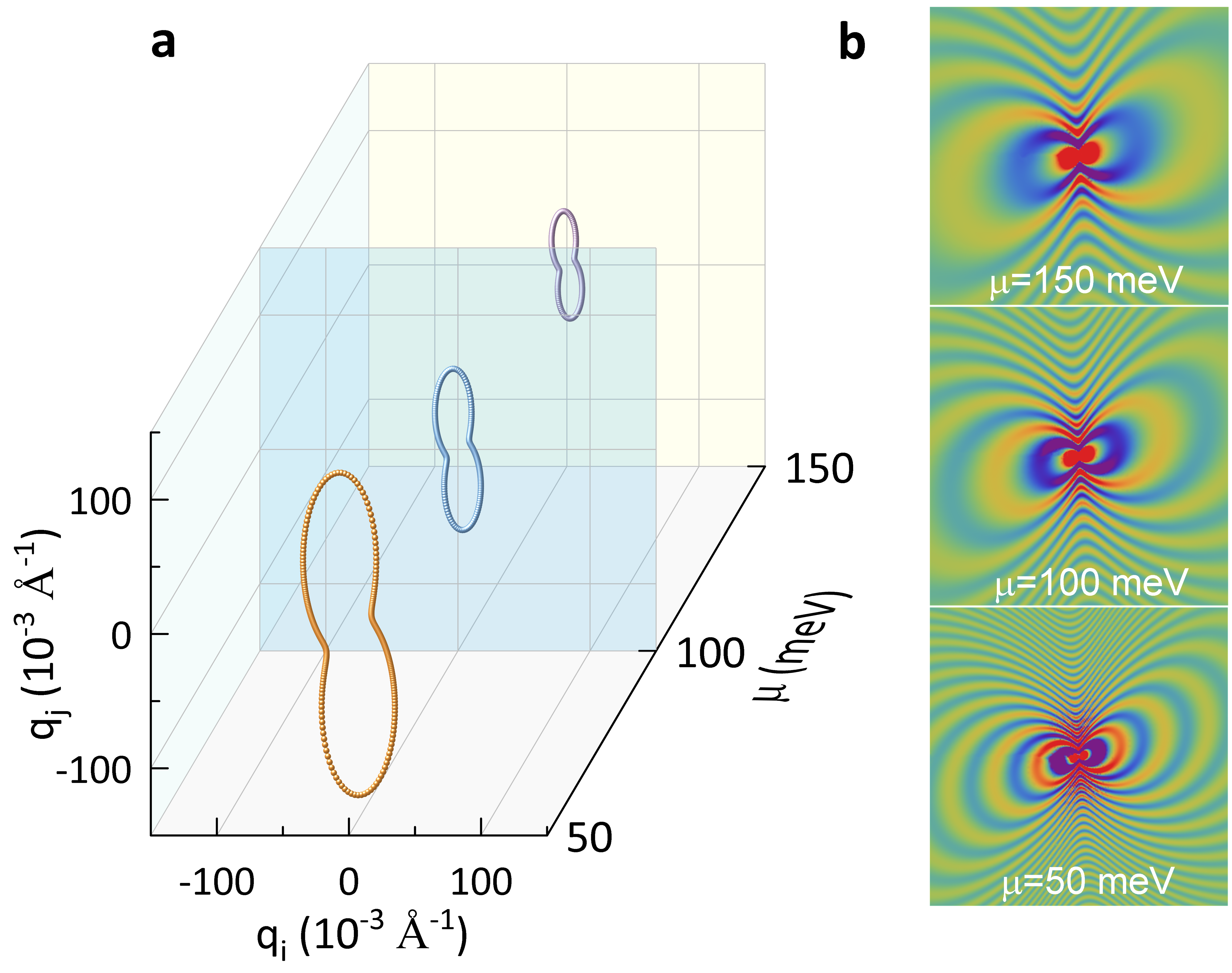}
\caption{\textbf{Tuning the wavelength of the chiral plasmon-polariton by gating/doping in a black phosphorous monolayer twisted with respect to its calcite substrate.} (a) Isofrequency contours of plasmon polaritons at fixed excitation energy $\hbar\omega = 170$ meV, for the chemical potentials $\mu = 50$ meV, $100$ meV, and $150$ meV, at a twist angle corresponding to the maximal chirality in Figure \ref{Fig: twist_isofreq} ($\phi = 47^\circ$). 
(b) The real-space profiles of the s-SNOM generated plasmon polaritons corresponding to those considered in panel (a).}
\label{twist_propagation}
\end{figure}

Finally, we discuss the influence of the chemical potential $\mu$ on the chiral propagation of plasmons, since chemical potential can be tuned in experiment by gating or doping~\cite{pogna2024bp}. Figure \ref{twist_propagation}(a) shows the isofrequencies for same plasmon excitation energy $\hbar \omega=170$ meV and the same twist angle $\phi=\phi_{c-max}$, but for varied chemical potential $\mu=50$ meV, $100$ meV and $150$ meV. Actually, the chemical potential does not affect the chirality, as all three contours in Figure \ref{twist_propagation}(a) preserve the identical shape and the same asymmetry ratio $\eta$. However, the chemical potential effectively tunes the plasmon wavelength in the host 2D material, as directly seen from Eq.~(\ref{Eq: dispersion_aniso}) and the changing momentum-scale of the isofrequency contours in Figure \ref{twist_propagation}(a).  
Figure \ref{twist_propagation}(b) shows the corresponding real-space propagation profiles. We note that for systems of finite size, highly directional propagation of plasmon polaritons can still be achieved even in the absence of closed isofrequency contours. This can take place when the plasmon wavelength ($\lambda_p\sim2\pi/q$) is much larger in one direction than in the other ($q_j\gg q_i$), so the plasmon is unable to propagate in the $i$-direction because of the limited system size, which can be tuned via doping or gating.

In conclusion, we have demonstrated how highly directional chiral plasmon polaritons can be generated in twisted photonic heterostructures through an intricate interplay between the anisotropic electronic structure of the material hosting the plasmons and the anisotropic electron-electron interactions fostered by the substrate. The chirality can be broadly tuned using not only the twist angle, but also the gating/doping as well as the excitation energy of the plasmons, all easily controllable in the readily existing experimental setups. This tuning of the plasmon polariton is far more versatile, broad, yet precise, than what would be feasible for phonon polaritons, because of the advantageous range of existing manipulations for electronic states in materials compared to those able to alter the lattice vibration frequencies. Our proposal therefore unveils a new pathway for controlled asymmetry and chiral propagation of plasmon polaritons at the nanoscale, readily achievable with existing experimental techniques for fabrication, excitation, manipulation and detection, and may thus immediately contribute to the development of compact yet tunable quantum optical devices for nanophotonics and on-chip optics.

\section{Acknowledgments}  
This work is supported by the Research Foundation-Flanders (FWO-Vlaanderen) and by the National Council for Scientific and Technological Development of Brazil (CNPq), through the PRONEX/FUNCAP, PQ (312705/2022-0), and UNIVERSAL (423423/2021-5) programs. Z. H. Tao gratefully acknowledges support from the China Scholarship Council.

\section{Supplementary Materials Available}
In the Supplementary Materials (SM), we provide the details of the calculations for the effective anisotropic electron-electron Coulomb interaction, and all the calculated isofrequency contours of the plasmon polaritons for various anisotropies of the 2D material and the substrate, as well as the twist angle between them.

\providecommand{\latin}[1]{#1}
\makeatletter
\providecommand{\doi}
  {\begingroup\let\do\@makeother\dospecials
  \catcode`\{=1 \catcode`\}=2 \doi@aux}
\providecommand{\doi@aux}[1]{\endgroup\texttt{#1}}
\makeatother
\providecommand*\mcitethebibliography{\thebibliography}
\csname @ifundefined\endcsname{endmcitethebibliography}
  {\let\endmcitethebibliography\endthebibliography}{}

\bibliography{ref}

\begin{mcitethebibliography}{32}
\providecommand*\natexlab[1]{#1}
\providecommand*\mciteSetBstSublistMode[1]{}
\providecommand*\mciteSetBstMaxWidthForm[2]{}
\providecommand*\mciteBstWouldAddEndPuncttrue
  {\def\EndOfBibitem{\unskip.}}
\providecommand*\mciteBstWouldAddEndPunctfalse
  {\let\EndOfBibitem\relax}
\providecommand*\mciteSetBstMidEndSepPunct[3]{}
\providecommand*\mciteSetBstSublistLabelBeginEnd[3]{}
\providecommand*\EndOfBibitem{}
\mciteSetBstSublistMode{f}
\mciteSetBstMaxWidthForm{subitem}{(\alph{mcitesubitemcount})}
\mciteSetBstSublistLabelBeginEnd
  {\mcitemaxwidthsubitemform\space}
  {\relax}
  {\relax}

\bibitem[Hu \latin{et~al.}(2023)Hu, Chen, Teng, Yu, Xue, Chen, Xiao, Qu, Hu,
  Chen, \latin{et~al.} others]{hu2023nanophotonic}
Hu,~H.; Chen,~N.; Teng,~H.; Yu,~R.; Xue,~M.; Chen,~K.; Xiao,~Y.; Qu,~Y.;
  Hu,~D.; Chen,~J., \latin{et~al.}  Gate-tunable negative refraction of
  mid-infrared polaritons. \emph{Science} \textbf{2023}, \emph{379},
  558--561\relax
\mciteBstWouldAddEndPuncttrue
\mciteSetBstMidEndSepPunct{\mcitedefaultmidpunct}
{\mcitedefaultendpunct}{\mcitedefaultseppunct}\relax
\EndOfBibitem
\bibitem[Oh \latin{et~al.}(2021)Oh, Altug, Jin, Low, Koester, Ivanov, Edel,
  Avouris, and Strano]{oh2021nanophotonic}
Oh,~S.~H.; Altug,~H.; Jin,~X.; Low,~T.; Koester,~S.~J.; Ivanov,~A.~P.;
  Edel,~J.~B.; Avouris,~P.; Strano,~M.~S. Nanophotonic biosensors harnessing
  van der Waals materials. \emph{Nature Communications} \textbf{2021},
  \emph{12}, 3824\relax
\mciteBstWouldAddEndPuncttrue
\mciteSetBstMidEndSepPunct{\mcitedefaultmidpunct}
{\mcitedefaultendpunct}{\mcitedefaultseppunct}\relax
\EndOfBibitem
\bibitem[Chaudhary \latin{et~al.}(2019)Chaudhary, Tamagnone, Yin, Sp{\"a}gele,
  Oscurato, Li, Persch, Li, Rubin, Jauregui, \latin{et~al.}
  others]{chaudhary2019nanophotonic}
Chaudhary,~K.; Tamagnone,~M.; Yin,~X.; Sp{\"a}gele,~C.~M.; Oscurato,~S.~L.;
  Li,~J.; Persch,~C.; Li,~R.; Rubin,~N.~A.; Jauregui,~L.~A., \latin{et~al.}
  Polariton nanophotonics using phase-change materials. \emph{Nature
  Communications} \textbf{2019}, \emph{10}, 4487\relax
\mciteBstWouldAddEndPuncttrue
\mciteSetBstMidEndSepPunct{\mcitedefaultmidpunct}
{\mcitedefaultendpunct}{\mcitedefaultseppunct}\relax
\EndOfBibitem
\bibitem[Wu \latin{et~al.}(2022)Wu, Duan, Ma, Ou, Li, Alonso-Gonz{\'a}lez,
  Caldwell, and Bao]{wu2022manipulating}
Wu,~Y.; Duan,~J.; Ma,~W.; Ou,~Q.; Li,~P.; Alonso-Gonz{\'a}lez,~P.;
  Caldwell,~J.~D.; Bao,~Q. Manipulating polaritons at the extreme scale in van
  der Waals materials. \emph{Nature Reviews Physics} \textbf{2022}, \emph{4},
  578--594\relax
\mciteBstWouldAddEndPuncttrue
\mciteSetBstMidEndSepPunct{\mcitedefaultmidpunct}
{\mcitedefaultendpunct}{\mcitedefaultseppunct}\relax
\EndOfBibitem
\bibitem[Zhang \latin{et~al.}(2021)Zhang, Hu, Ma, Li, Krasnok, Hillenbrand,
  Al{\`u}, and Qiu]{zhang2021interface}
Zhang,~Q.; Hu,~G.; Ma,~W.; Li,~P.; Krasnok,~A.; Hillenbrand,~R.; Al{\`u},~A.;
  Qiu,~C.~W. Interface nano-optics with van der Waals polaritons. \emph{Nature}
  \textbf{2021}, \emph{597}, 187--195\relax
\mciteBstWouldAddEndPuncttrue
\mciteSetBstMidEndSepPunct{\mcitedefaultmidpunct}
{\mcitedefaultendpunct}{\mcitedefaultseppunct}\relax
\EndOfBibitem
\bibitem[Liu \latin{et~al.}(2016)Liu, Li, Sadana, and Sorger]{liu2016onchip}
Liu,~K.; Li,~N.; Sadana,~D.~K.; Sorger,~V.~J. Integrated nanocavity plasmon
  light sources for on-chip optical interconnects. \emph{ACS Photonics}
  \textbf{2016}, \emph{3}, 233--242\relax
\mciteBstWouldAddEndPuncttrue
\mciteSetBstMidEndSepPunct{\mcitedefaultmidpunct}
{\mcitedefaultendpunct}{\mcitedefaultseppunct}\relax
\EndOfBibitem
\bibitem[Schwarz \latin{et~al.}(2014)Schwarz, Reininger, Ristani{\'c}, Detz,
  Andrews, Schrenk, and Strasser]{schwarz2014monolithically}
Schwarz,~B.; Reininger,~P.; Ristani{\'c},~D.; Detz,~H.; Andrews,~A.~M.;
  Schrenk,~W.; Strasser,~G. Monolithically integrated mid-infrared
  lab-on-a-chip using plasmonics and quantum cascade structures. \emph{Nature
  Communications} \textbf{2014}, \emph{5}, 4085\relax
\mciteBstWouldAddEndPuncttrue
\mciteSetBstMidEndSepPunct{\mcitedefaultmidpunct}
{\mcitedefaultendpunct}{\mcitedefaultseppunct}\relax
\EndOfBibitem
\bibitem[Fu \latin{et~al.}(2024)Fu, Qu, Xue, Liu, Chen, Zhao, Chen, Li, Weng,
  Liu, Dai, and Chen]{fu_manipulating_2024}
Fu,~R.; Qu,~Y.; Xue,~M.; Liu,~X.; Chen,~S.; Zhao,~Y.; Chen,~R.; Li,~B.;
  Weng,~H.; Liu,~Q.; Dai,~Q.; Chen,~J. Manipulating hyperbolic transient
  plasmons in a layered semiconductor. \emph{Nature Communications}
  \textbf{2024}, \emph{15}, 709\relax
\mciteBstWouldAddEndPuncttrue
\mciteSetBstMidEndSepPunct{\mcitedefaultmidpunct}
{\mcitedefaultendpunct}{\mcitedefaultseppunct}\relax
\EndOfBibitem
\bibitem[Low \latin{et~al.}(2014)Low, Rold{\'a}n, Wang, Xia, Avouris, Moreno,
  and Guinea]{low2014plasmons}
Low,~T.; Rold{\'a}n,~R.; Wang,~H.; Xia,~F.; Avouris,~P.; Moreno,~L.~M.;
  Guinea,~F. Plasmons and screening in monolayer and multilayer black
  phosphorus. \emph{Physical Review Letters} \textbf{2014}, \emph{113},
  106802\relax
\mciteBstWouldAddEndPuncttrue
\mciteSetBstMidEndSepPunct{\mcitedefaultmidpunct}
{\mcitedefaultendpunct}{\mcitedefaultseppunct}\relax
\EndOfBibitem
\bibitem[Pogna \latin{et~al.}(2024)Pogna, Pistore, Viti, Li, Davies, Linfield,
  and Vitiello]{pogna2024bp}
Pogna,~E.~A.; Pistore,~V.; Viti,~L.; Li,~L.; Davies,~A.~G.; Linfield,~E.~H.;
  Vitiello,~M.~S. Near-field detection of gate-tunable anisotropic plasmon
  polaritons in black phosphorus at terahertz frequencies. \emph{Nature
  Communications} \textbf{2024}, \emph{15}, 2373\relax
\mciteBstWouldAddEndPuncttrue
\mciteSetBstMidEndSepPunct{\mcitedefaultmidpunct}
{\mcitedefaultendpunct}{\mcitedefaultseppunct}\relax
\EndOfBibitem
\bibitem[Galiffi \latin{et~al.}(2024)Galiffi, Carini, Ni,
  {\'A}lvarez-P{\'e}rez, Yves, Renzi, Nolen, Wasserroth, Wolf, Alonso-Gonzalez,
  \latin{et~al.} others]{galiffi2023review_aniso_PhPs}
Galiffi,~E.; Carini,~G.; Ni,~X.; {\'A}lvarez-P{\'e}rez,~G.; Yves,~S.;
  Renzi,~E.~M.; Nolen,~R.; Wasserroth,~S.; Wolf,~M.; Alonso-Gonzalez,~P.,
  \latin{et~al.}  Extreme light confinement and control in low-symmetry
  phonon-polaritonic crystals. \emph{Nature Reviews Materials} \textbf{2024},
  \emph{9}, 9--28\relax
\mciteBstWouldAddEndPuncttrue
\mciteSetBstMidEndSepPunct{\mcitedefaultmidpunct}
{\mcitedefaultendpunct}{\mcitedefaultseppunct}\relax
\EndOfBibitem
\bibitem[Passler \latin{et~al.}(2022)Passler, Ni, Hu, Matson, Carini, Wolf,
  Schubert, Al{\`u}, Caldwell, Folland, \latin{et~al.}
  others]{passler2022shear}
Passler,~N.~C.; Ni,~X.; Hu,~G.; Matson,~J.~R.; Carini,~G.; Wolf,~M.;
  Schubert,~M.; Al{\`u},~A.; Caldwell,~J.~D.; Folland,~T.~G., \latin{et~al.}
  Hyperbolic shear polaritons in low-symmetry crystals. \emph{Nature}
  \textbf{2022}, \emph{602}, 595--600\relax
\mciteBstWouldAddEndPuncttrue
\mciteSetBstMidEndSepPunct{\mcitedefaultmidpunct}
{\mcitedefaultendpunct}{\mcitedefaultseppunct}\relax
\EndOfBibitem
\bibitem[Hu \latin{et~al.}(2023)Hu, Ma, Hu, Wu, Zheng, Liu, Zhang, Ni, Chen,
  Zhang, \latin{et~al.} others]{hu2023shear}
Hu,~G.; Ma,~W.; Hu,~D.; Wu,~J.; Zheng,~C.; Liu,~K.; Zhang,~X.; Ni,~X.;
  Chen,~J.; Zhang,~X., \latin{et~al.}  Real-space nanoimaging of hyperbolic
  shear polaritons in a monoclinic crystal. \emph{Nature Nanotechnology}
  \textbf{2023}, \emph{18}, 64--70\relax
\mciteBstWouldAddEndPuncttrue
\mciteSetBstMidEndSepPunct{\mcitedefaultmidpunct}
{\mcitedefaultendpunct}{\mcitedefaultseppunct}\relax
\EndOfBibitem
\bibitem[Matson \latin{et~al.}(2023)Matson, Wasserroth, Ni, Obst,
  Diaz-Granados, Carini, Renzi, Galiffi, Folland, Eng, \latin{et~al.}
  others]{matson2023shear}
Matson,~J.; Wasserroth,~S.; Ni,~X.; Obst,~M.; Diaz-Granados,~K.; Carini,~G.;
  Renzi,~E.~M.; Galiffi,~E.; Folland,~T.~G.; Eng,~L.~M., \latin{et~al.}
  Controlling the propagation asymmetry of hyperbolic shear polaritons in
  beta-gallium oxide. \emph{Nature Communications} \textbf{2023}, \emph{14},
  5240\relax
\mciteBstWouldAddEndPuncttrue
\mciteSetBstMidEndSepPunct{\mcitedefaultmidpunct}
{\mcitedefaultendpunct}{\mcitedefaultseppunct}\relax
\EndOfBibitem
\bibitem[{\'A}lvarez-Cuervo \latin{et~al.}(2024){\'A}lvarez-Cuervo, Obst,
  Dixit, Carini, Tresguerres-Mata, Lanza, Ter{\'a}n-Garc{\'\i}a,
  {\'A}lvarez-P{\'e}rez, Fern{\'a}ndez-{\'A}lvarez, Diaz-Granados,
  \latin{et~al.} others]{alvarez2024unidirectional}
{\'A}lvarez-Cuervo,~J.; Obst,~M.; Dixit,~S.; Carini,~G.; Tresguerres-Mata,~A.;
  Lanza,~C.; Ter{\'a}n-Garc{\'\i}a,~E.; {\'A}lvarez-P{\'e}rez,~G.;
  Fern{\'a}ndez-{\'A}lvarez,~L.; Diaz-Granados,~K., \latin{et~al.}
  Unidirectional Ray Polaritons in Twisted Asymmetric Stacks. \emph{arXiv
  preprint arXiv:2403.18657} \textbf{2024}, \relax
\mciteBstWouldAddEndPunctfalse
\mciteSetBstMidEndSepPunct{\mcitedefaultmidpunct}
{}{\mcitedefaultseppunct}\relax
\EndOfBibitem
\bibitem[Low and Avouris(2014)Low, and Avouris]{low2014graphene}
Low,~T.; Avouris,~P. Graphene plasmonics for terahertz to mid-infrared
  applications. \emph{ACS Nano} \textbf{2014}, \emph{8}, 1086--1101\relax
\mciteBstWouldAddEndPuncttrue
\mciteSetBstMidEndSepPunct{\mcitedefaultmidpunct}
{\mcitedefaultendpunct}{\mcitedefaultseppunct}\relax
\EndOfBibitem
\bibitem[Lundeberg \latin{et~al.}(2017)Lundeberg, Gao, Asgari, Tan, Van~Duppen,
  Autore, Alonso-Gonz{\'a}lez, Woessner, Watanabe, Taniguchi, \latin{et~al.}
  others]{lundeberg2017tuning}
Lundeberg,~M.~B.; Gao,~Y.; Asgari,~R.; Tan,~C.; Van~Duppen,~B.; Autore,~M.;
  Alonso-Gonz{\'a}lez,~P.; Woessner,~A.; Watanabe,~K.; Taniguchi,~T.,
  \latin{et~al.}  Tuning quantum nonlocal effects in graphene plasmonics.
  \emph{Science} \textbf{2017}, \emph{357}, 187--191\relax
\mciteBstWouldAddEndPuncttrue
\mciteSetBstMidEndSepPunct{\mcitedefaultmidpunct}
{\mcitedefaultendpunct}{\mcitedefaultseppunct}\relax
\EndOfBibitem
\bibitem[Hossain \latin{et~al.}(2009)Hossain, Murch, Belova, and
  Turner]{hossain2009electronic}
Hossain,~F.~M.; Murch,~G.~E.; Belova,~I.~V.; Turner,~B.~D. Electronic, optical
  and bonding properties of CaCO3 calcite. \emph{Solid State Communications}
  \textbf{2009}, \emph{149}, 1201--1203\relax
\mciteBstWouldAddEndPuncttrue
\mciteSetBstMidEndSepPunct{\mcitedefaultmidpunct}
{\mcitedefaultendpunct}{\mcitedefaultseppunct}\relax
\EndOfBibitem
\bibitem[Huang \latin{et~al.}(2022)Huang, Tu, Shen, Zheng, Wang, Wang, Khaliji,
  Park, Liu, Yang, \latin{et~al.} others]{huang2022chiralplasmons}
Huang,~T.; Tu,~X.; Shen,~C.; Zheng,~B.; Wang,~J.; Wang,~H.; Khaliji,~K.;
  Park,~S.~H.; Liu,~Z.; Yang,~T., \latin{et~al.}  Observation of chiral and
  slow plasmons in twisted bilayer graphene. \emph{Nature} \textbf{2022},
  \emph{605}, 63--68\relax
\mciteBstWouldAddEndPuncttrue
\mciteSetBstMidEndSepPunct{\mcitedefaultmidpunct}
{\mcitedefaultendpunct}{\mcitedefaultseppunct}\relax
\EndOfBibitem
\bibitem[Qiao \latin{et~al.}(2014)Qiao, Kong, Hu, Yang, and Ji]{qiao_2014}
Qiao,~J.~S.; Kong,~X.~H.; Hu,~Z.~X.; Yang,~F.; Ji,~W. High-mobility transport
  anisotropy and linear dichroism in few-layer black phosphorus. \emph{Nature
  Communications} \textbf{2014}, \emph{5}, 4475\relax
\mciteBstWouldAddEndPuncttrue
\mciteSetBstMidEndSepPunct{\mcitedefaultmidpunct}
{\mcitedefaultendpunct}{\mcitedefaultseppunct}\relax
\EndOfBibitem
\bibitem[Ghosh \latin{et~al.}(2017)Ghosh, Kumar, Thakur, Chauhan, Bhowmick, and
  Agarwal]{ghosh2017anisotropic}
Ghosh,~B.; Kumar,~P.; Thakur,~A.; Chauhan,~Y.~S.; Bhowmick,~S.; Agarwal,~A.
  Anisotropic plasmons, excitons, and electron energy loss spectroscopy of
  phosphorene. \emph{Physical Review B} \textbf{2017}, \emph{96}, 035422\relax
\mciteBstWouldAddEndPuncttrue
\mciteSetBstMidEndSepPunct{\mcitedefaultmidpunct}
{\mcitedefaultendpunct}{\mcitedefaultseppunct}\relax
\EndOfBibitem
\bibitem[Giuliani and Vignale(2008)Giuliani, and Vignale]{giuliani2008quantum}
Giuliani,~G.; Vignale,~G. \emph{Quantum theory of the electron liquid};
  Cambridge university press, 2008\relax
\mciteBstWouldAddEndPuncttrue
\mciteSetBstMidEndSepPunct{\mcitedefaultmidpunct}
{\mcitedefaultendpunct}{\mcitedefaultseppunct}\relax
\EndOfBibitem
\bibitem[Dong \latin{et~al.}(2019)Dong, Tao, Duan, Huang, and Zhao]{Dong_2020}
Dong,~H.~M.; Tao,~Z.~H.; Duan,~Y.~F.; Huang,~F.; Zhao,~C.~X. Coupled
  plasmon-phonon modes in monolayer MoS$_2$. \emph{Journal of Physics:
  Condensed Matter} \textbf{2019}, \emph{32}, 125703\relax
\mciteBstWouldAddEndPuncttrue
\mciteSetBstMidEndSepPunct{\mcitedefaultmidpunct}
{\mcitedefaultendpunct}{\mcitedefaultseppunct}\relax
\EndOfBibitem
\bibitem[Rodin and Castro~Neto(2015)Rodin, and
  Castro~Neto]{Rodin2015Collective}
Rodin,~A.~S.; Castro~Neto,~A.~H. Collective modes in anisotropic double-layer
  systems. \emph{Physical Review B} \textbf{2015}, \emph{91}, 075422\relax
\mciteBstWouldAddEndPuncttrue
\mciteSetBstMidEndSepPunct{\mcitedefaultmidpunct}
{\mcitedefaultendpunct}{\mcitedefaultseppunct}\relax
\EndOfBibitem
\bibitem[Pyatkovskiy and Chakraborty(2016)Pyatkovskiy, and
  Chakraborty]{pyatkovskiy2016dynamical}
Pyatkovskiy,~P.; Chakraborty,~T. Dynamical polarization and plasmons in a
  two-dimensional system with merging Dirac points. \emph{Physical Review B}
  \textbf{2016}, \emph{93}, 085145\relax
\mciteBstWouldAddEndPuncttrue
\mciteSetBstMidEndSepPunct{\mcitedefaultmidpunct}
{\mcitedefaultendpunct}{\mcitedefaultseppunct}\relax
\EndOfBibitem
\bibitem[Silva-Guill{\'e}n \latin{et~al.}(2017)Silva-Guill{\'e}n, Canadell,
  Ordej{\'o}n, Guinea, and Rold{\'a}n]{silva2017anisotropic}
Silva-Guill{\'e}n,~J.~A.; Canadell,~E.; Ordej{\'o}n,~P.; Guinea,~F.;
  Rold{\'a}n,~R. Anisotropic features in the electronic structure of the
  two-dimensional transition metal trichalcogenide TiS3: electron doping and
  plasmons. \emph{2D Materials} \textbf{2017}, \emph{4}, 025085\relax
\mciteBstWouldAddEndPuncttrue
\mciteSetBstMidEndSepPunct{\mcitedefaultmidpunct}
{\mcitedefaultendpunct}{\mcitedefaultseppunct}\relax
\EndOfBibitem
\bibitem[Zhu \latin{et~al.}(2018)Zhu, Yi, Li, Xiao, Zhang, Yang, Kaindl, Li,
  Wang, and Zhang]{zhu2018chiralphonons}
Zhu,~H.; Yi,~J.; Li,~M.~Y.; Xiao,~J.; Zhang,~L.; Yang,~C.~W.; Kaindl,~R.~A.;
  Li,~L.~J.; Wang,~Y.; Zhang,~X. Observation of chiral phonons. \emph{Science}
  \textbf{2018}, \emph{359}, 579--582\relax
\mciteBstWouldAddEndPuncttrue
\mciteSetBstMidEndSepPunct{\mcitedefaultmidpunct}
{\mcitedefaultendpunct}{\mcitedefaultseppunct}\relax
\EndOfBibitem
\bibitem[Ma \latin{et~al.}(2021)Ma, Hu, Hu, Chen, Sun, Zhang, Dai, Zeng,
  Al{\`u}, Qiu, \latin{et~al.} others]{ma2021ghost}
Ma,~W.; Hu,~G.; Hu,~D.; Chen,~R.~K.; Sun,~T.; Zhang,~X.; Dai,~Q.; Zeng,~Y.;
  Al{\`u},~A.; Qiu,~C.~W., \latin{et~al.}  Ghost hyperbolic surface polaritons
  in bulk anisotropic crystals. \emph{Nature} \textbf{2021}, \emph{596},
  362--366\relax
\mciteBstWouldAddEndPuncttrue
\mciteSetBstMidEndSepPunct{\mcitedefaultmidpunct}
{\mcitedefaultendpunct}{\mcitedefaultseppunct}\relax
\EndOfBibitem
\bibitem[Woessner \latin{et~al.}(2015)Woessner, Lundeberg, Gao, Principi,
  Alonso-Gonz{\'a}lez, Carrega, Watanabe, Taniguchi, Vignale, Polini,
  \latin{et~al.} others]{woessner2015ssnom}
Woessner,~A.; Lundeberg,~M.~B.; Gao,~Y.; Principi,~A.; Alonso-Gonz{\'a}lez,~P.;
  Carrega,~M.; Watanabe,~K.; Taniguchi,~T.; Vignale,~G.; Polini,~M.,
  \latin{et~al.}  Highly confined low-loss plasmons in graphene--boron nitride
  heterostructures. \emph{Nature Materials} \textbf{2015}, \emph{14},
  421--425\relax
\mciteBstWouldAddEndPuncttrue
\mciteSetBstMidEndSepPunct{\mcitedefaultmidpunct}
{\mcitedefaultendpunct}{\mcitedefaultseppunct}\relax
\EndOfBibitem
\bibitem[Ma \latin{et~al.}(2018)Ma, Alonso-Gonz{\'a}lez, Li, Nikitin, Yuan,
  Mart{\'\i}n-S{\'a}nchez, Taboada-Guti{\'e}rrez, Amenabar, Li, V{\'e}lez,
  \latin{et~al.} others]{ma2018hPhPs}
Ma,~W.; Alonso-Gonz{\'a}lez,~P.; Li,~S.; Nikitin,~A.~Y.; Yuan,~J.;
  Mart{\'\i}n-S{\'a}nchez,~J.; Taboada-Guti{\'e}rrez,~J.; Amenabar,~I.; Li,~P.;
  V{\'e}lez,~S., \latin{et~al.}  In-plane anisotropic and ultra-low-loss
  polaritons in a natural van der Waals crystal. \emph{Nature} \textbf{2018},
  \emph{562}, 557--562\relax
\mciteBstWouldAddEndPuncttrue
\mciteSetBstMidEndSepPunct{\mcitedefaultmidpunct}
{\mcitedefaultendpunct}{\mcitedefaultseppunct}\relax
\EndOfBibitem
\bibitem[Tao \latin{et~al.}(2021)Tao, Dong, Milo{\v{s}}evi{\'c}, Peeters, and
  Van~Duppen]{tao2021tailoring}
Tao,~Z.~H.; Dong,~H.~M.; Milo{\v{s}}evi{\'c},~M.; Peeters,~F.~M.;
  Van~Duppen,~B. Tailoring Dirac plasmons via anisotropic dielectric
  environment by design. \emph{Physical Review Applied} \textbf{2021},
  \emph{16}, 054030\relax
\mciteBstWouldAddEndPuncttrue
\mciteSetBstMidEndSepPunct{\mcitedefaultmidpunct}
{\mcitedefaultendpunct}{\mcitedefaultseppunct}\relax
\EndOfBibitem
\end{mcitethebibliography}
\end{document}


\section{I. The effective anisotropic electron-electron Coulomb interaction}

\begin{figure}[H]
\includegraphics[width=0.6\textwidth]{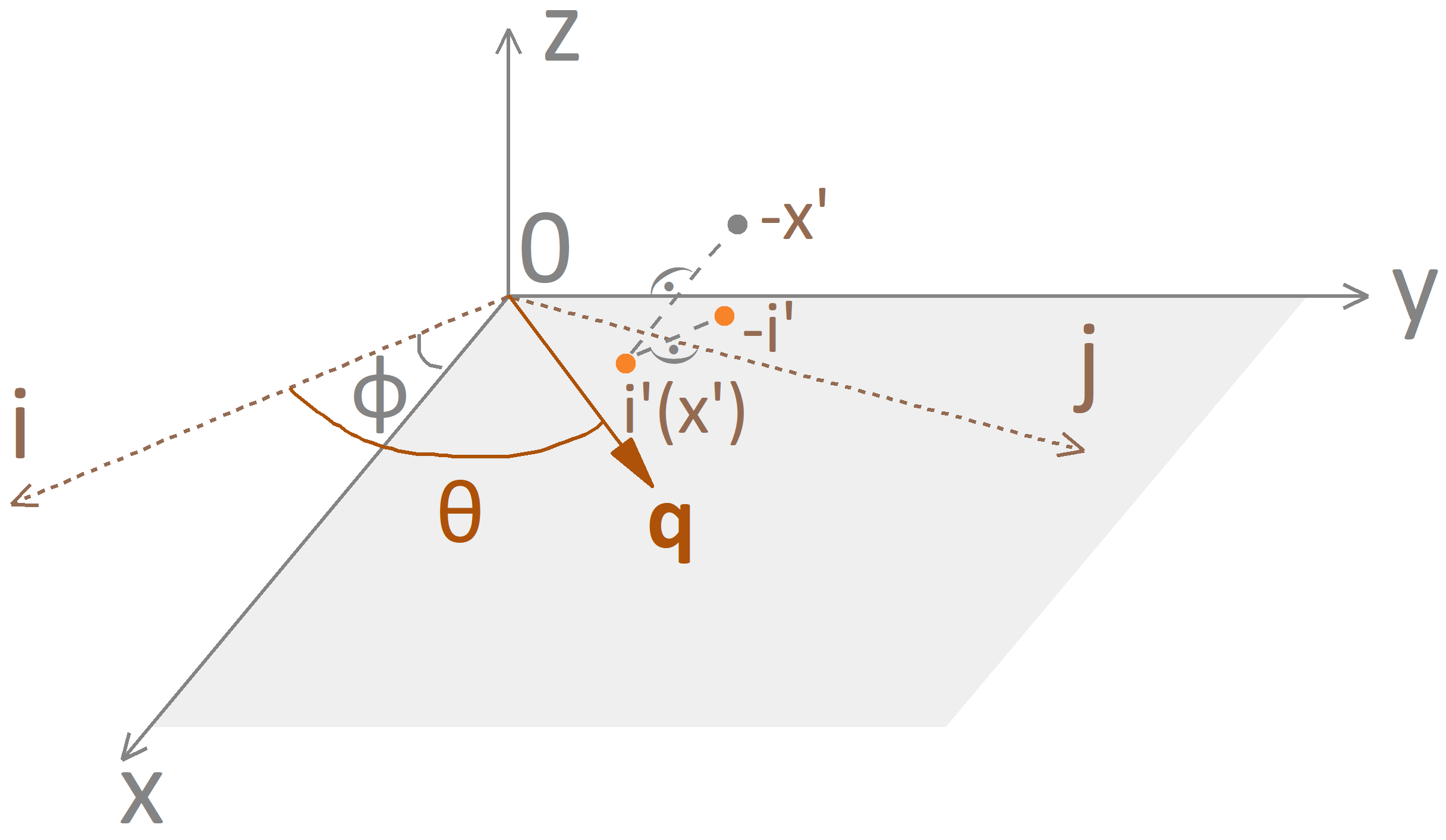}
\caption{Alignment of the $x$-$y$ coordinate axes (gray solid arrows) of the 2D material, twisted by $\phi$ relative to the $i$-$j$ axes (brown dot arrows) of the substrate. }
\label{S1}
\end{figure}

The dielectric tensor of the environment above and below the 2D material layer ($z=0$ plane) is expressed as follows
\begin{equation}
\tilde{\varepsilon}_s=\left[\begin{array}{ccc}
\varepsilon_{s,x} & 0 & 0 \\
0 & \varepsilon_{s,y} & 0 \\
0 & 0 & \varepsilon_{s,z}
\tag{S1}
\end{array}\right].
\end{equation}
The electric field $\boldsymbol{E}_s(\boldsymbol{r},z)=-{ \boldsymbol\nabla}\Phi_s(\boldsymbol{r},z)$ and displacement field ${\boldsymbol D}_s(\boldsymbol{r},z)=\tilde{\varepsilon}_s\cdot\boldsymbol{E}_s(\boldsymbol{r},z)$, with $\boldsymbol{\nabla}\cdot{\boldsymbol D}_s(\boldsymbol{r},z) = 0$, are determined from the  electric potential $\Phi_s(\boldsymbol{r},z)$.  In the cylindrical coordinate system, $\boldsymbol{r}=(x,y)$ is a 2D vector in the $z = 0$ plane. The calculations are divided into an upper region $z > 0$ and a lower region $z < 0$, indexed by $s=\pm$. 

The anisotropic electron-electron Coulomb interaction is obtained by solving the electrostatic problem \cite{PhysRevB165408}, 
\begin{equation}
[\varepsilon_{s,x}\partial_x^2+\varepsilon_{s,y}\partial_y^2+\varepsilon_{s,z}\partial_z^2]
\Phi_s({\boldsymbol r}, z)=0 \ ,
\tag{S2}
\label{eq:electrostatic}
\end{equation}
with 3D electrical potential of amplitude $\varphi_s$,
\begin{equation}
    \Phi_s(\boldsymbol{r},z)=\varphi_s e^{i\boldsymbol{q}_s\cdot\boldsymbol{r}-q_{z,s}|z|},
    \tag{S3}
    \label{eq:potential}
\end{equation}
and momenta, $\boldsymbol{q}_s=q_s[{\rm cos}(\theta-\phi),{\rm sin}(\theta-\phi)]$ and $q_{z,s}$, that are 
 wavevectors  in region $s$ respectively parallel and perpendicular to the $x-y$ plane. $\theta$ is the angle between $\boldsymbol{q}_s$ and the $i$-axis, and $\phi$ is the twist angle. Because of  
 the exponential factor $e^{-q_{z,s}|z|}$, the potential does not diverge when $z \rightarrow\pm\infty$.  
 
Substituting Eq.~(\ref{eq:potential}) into Eq.~(\ref{eq:electrostatic}), we obtain a relation between the in-plane and out-of-plane wave vectors for $z \neq 0$, 
\begin{equation}
q_{z,s}=q_s[(\varepsilon_{s,x}/\varepsilon_{s,z}){\rm cos}^2(\theta-\phi)+(\varepsilon_{s,y}/\varepsilon_{s,z}){\rm sin}^2(\theta-\phi)]^{1/2}.  
\tag{S4}
\end{equation}
The 2D layer with free electrons at the interface $z = 0$ can be
considered as a metallic boundary condition with
\begin{equation}
\hat{z}\times[\boldsymbol{E}_+-\boldsymbol{E}_-]=0,~
\hat{z}\cdot[\tilde{\boldsymbol{\varepsilon}}_+\cdot\boldsymbol{E}_+-\tilde{\boldsymbol{\varepsilon}}_-\cdot\boldsymbol{E}_-]=-4\pi e \cdot e^{i\boldsymbol{q}\cdot\boldsymbol{r}}.
\tag{S5}
\label{eq: boundary condition}
\end{equation}
Here $-e$ is the electron charge and $\boldsymbol{q}$  the in-plane wavevector for $z = 0$. Imposing for $z = 0$, continuity in the electrical potential, $\varphi_+|_{z=0^+} = \varphi_-|_{z=0^-} = \varphi$,  and conservation of momentum, $\boldsymbol{q}_+|_{z=0^+} = \boldsymbol{q}_-|_{z=0^-} = \boldsymbol{q}$, we obtain the effective electron-electron interaction,  
\begin{equation}
    \boldsymbol{V}(q,\theta,\phi)=-e\varphi=\frac{4\pi e^2}{\varepsilon_{\rm eff}(\theta,\phi)q}\ .
    \tag{S6}
    \label{eq: e-e interaction}
\end{equation}
The directional-dependent effective dielectric function reduces to,  
\begin{equation}
\varepsilon_{\rm eff}(\theta,\phi)=\sum\limits_{s=\pm}[(\varepsilon_{s,x}\varepsilon_{s,z}){\rm cos}^2(\theta-\phi)+(\varepsilon_{s,y}\varepsilon_{s,z}){\rm sin}^2(\theta-\phi)]^{1/2}.  
\tag{S7}
\end{equation}
Here we take the $s = + $ region as vacuum and $s = - $ region as a calcite substrate. Thus, taking into account the dielectric properties of this material, the dielectric constants in our system are considered as follows:
\begin{equation}
\varepsilon_{+,x}=\varepsilon_{+,y}=\varepsilon_{+,z}=\varepsilon_0=1,
\tag{S8}
\end{equation}
\begin{equation}
\varepsilon_{-,y}=\varepsilon_\perp=\varepsilon_{\infty,1}\left(1+\frac{\omega^2_{LO,1}-\omega^2_{TO,1}}{\omega^2_{TO,1}-\omega^2-i\omega\Gamma_1}+\frac{\omega^2_{LO,2}-\omega^2_{TO,2}}{\omega^2_{TO,2}-\omega^2-i\omega\Gamma_2}\right),
\tag{S9}
\label{aniso_epsilon_perp}
\end{equation}
\begin{equation}
\varepsilon_{-,x}=\varepsilon_{-,z}=\varepsilon_\parallel=\varepsilon_{\infty,3}\left(1+\frac{\omega^2_{LO,3}-\omega^2_{TO,3}}{\omega^2_{TO,3}-\omega^2-i\omega\Gamma_3}\right)\ , 
\tag{S10}
\label{aniso_epsilon_parallel}
\end{equation}
where $\varepsilon_{\infty,1} = 2.7$, $\omega^2_{TO,1} = 712$ cm$^{-1}$, $\omega^2_{LO,1} = 715$ cm$^{-1}$, $\Gamma_1 = 5$ cm$^{-1}$. $\omega^2_{TO,2} = 1410$ cm$^{-1}$, $\omega^2_{LO,2} = 1550$ cm$^{-1}$, $\Gamma_2 = 10$ cm$^{-1}$. $\varepsilon_{\infty,3} = 2.4$, $\omega^2_{TO,3} = 871$ cm$^{-1}$, $\omega^2_{LO,3} = 890$ cm$^{-1}$, $\Gamma_1 = 3$ cm$^{-1}$~\cite{ma2021ghost}.

\newpage

\section{II. Isofrequency contours of the plasmon polaritons in the monolayer black phosphorus on top of calcite}

Figure S2 shows the isofrequency contours of the plasmon polariton in a monolayer black phosphorus on top of a calcite substrate, as a function of the twist angle $\phi$ between the two principal anisotropy directions in the system.  

\begin{figure}[H]
\centering{}\includegraphics[width=0.95\columnwidth]{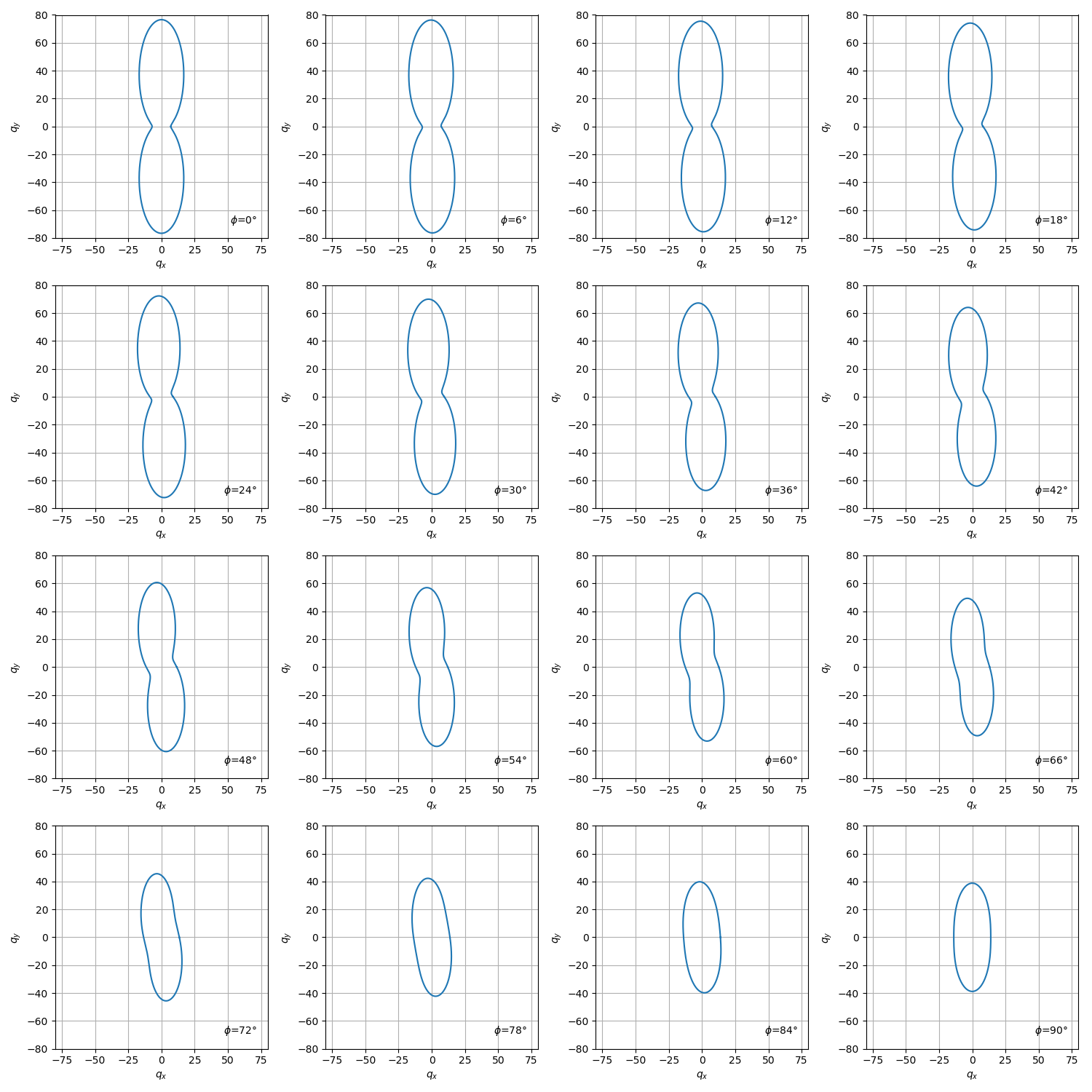}
\caption{Isofrequency contours of the plasmon polaritons at fixed excitation energy $\hbar\omega = 170$ meV and the fixed chemical potential $\mu = 100$ meV for twist angle $\phi$, as labelled. Momentum scale $q_{x,y}$ is in units of $10^{-3}$ $\AA^{-1}$.}
\label{S2}
\end{figure}

\newpage

\section{III. Isofrequency contours considering different anisotropies of the 2D material and the substrate}

\begin{figure}[H]
\centering{}\includegraphics[width=0.85\columnwidth]{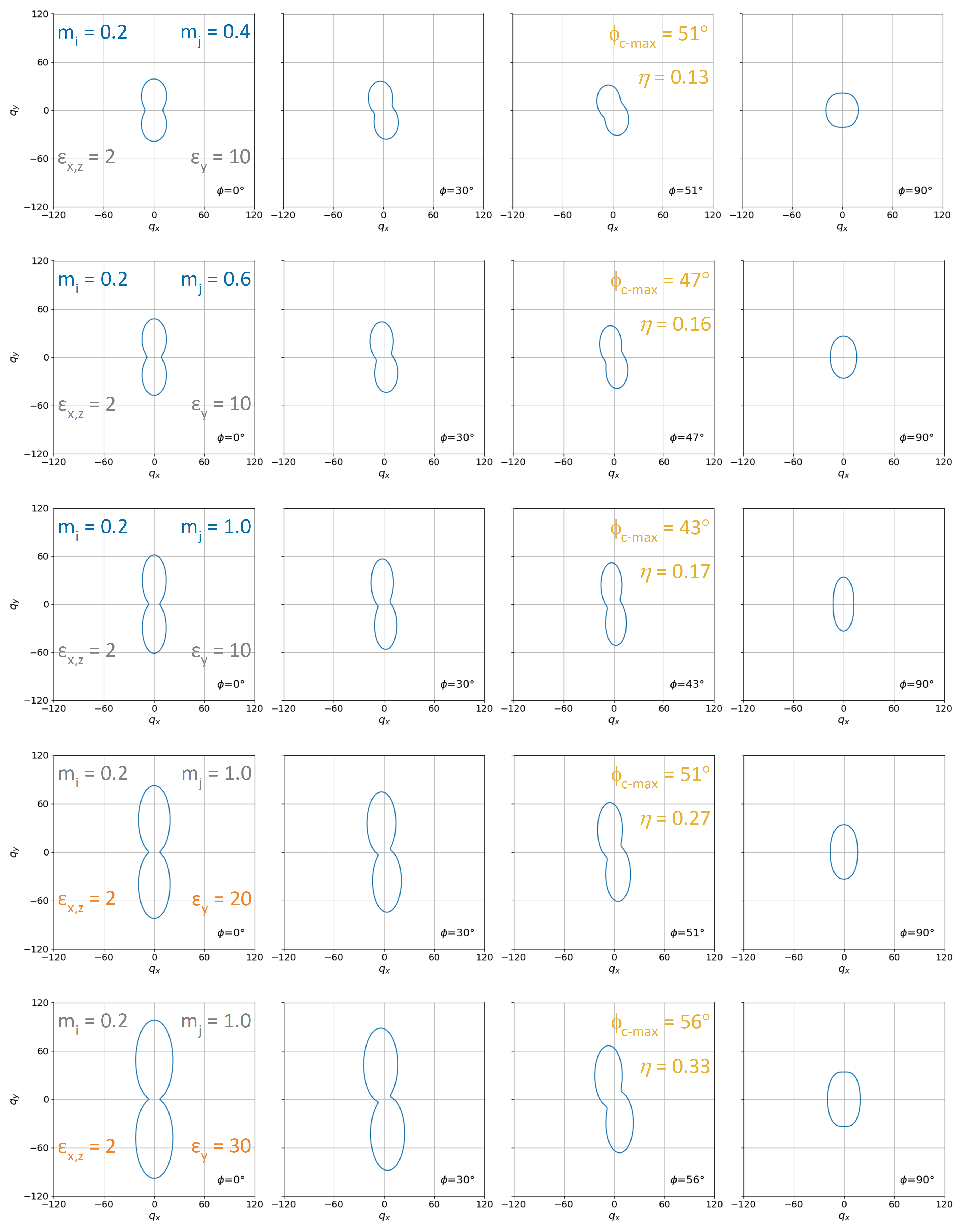}
\caption{Isofrequency contours of the plasmon polaritons at fixed excitation energy $\hbar\omega = 170$ meV and the fixed chemical potential $\mu = 100$ meV, for varied twist angles $\phi$ and anisotropies of the 2D material and the substrate, as labelled. Here, $\phi_{c-max}$ represents the twist angle yielding the maximal chirality, and $\eta$ denotes the corresponding asymmetry ratio. Momentum scale $q_{x,y}$ is in units of $10^{-3}$ $\AA^{-1}$.}
\label{S3}
\end{figure}